\documentclass{elsart}

\usepackage{graphicx}
\usepackage{amssymb}

\def\figurename#1#2{{\bf Figure~\thefigure}}

\def\PCMO{\rm {Pr_{1-{\it x}} Ca_{\it x} Mn O_3}}
\def\LCMO{\rm {La_{1-{\it x}} Ca_{\it x} Mn O_3}}
\def\BCMO{\rm {Bi_{1-{\it x}} Ca_{\it x} Mn O_3}}
\def\LPCMO{\rm {(La_{1-{\it y}} Pr_{\it y})_{1-{\it x}} Ca_{\it x} Mn O_3}}
\def\LSCO{\rm {La_{2-{\it x}} Sr_{\it x} Cu O_4}}
\def\TC{T_{\rm C}}
\def\Tc{T_{\rm c}}
\def\eg{e_{\rm g}}
\def\t2g{t_{\rm 2g}}
\def\JH{J_{\rm H}}
\def\JAF{J_{\rm AF}}

\begin{document}

\begin{frontmatter}
\title{NANOSCALE PHASE SEPARATION IN \\
COLOSSAL MAGNETORESISTANCE MATERIALS:\\
LESSONS FOR THE CUPRATES?\thanksref{proc}}

\thanks[proc]{To appear in the Proceedings of the
post-$\mu$SR2002 \emph{Superconductivity Workshop},
Williamsburg, VA., June 3-7, 2002, edited by
A.J. Millis, S. Uchida, Y.J. Uemura.}

\author{Elbio Dagotto$^1$, Jan Burgy$^2$ and Adriana Moreo$^3$}
\address{National High Magnetic Field Lab and Department of Physics,
Florida State University, Tallahassee, FL 32306, USA}

\begin{abstract}
A recent vast experimental and theoretical effort in manganites
has shown that the colossal magnetoresistance effect can be understood
based on the competition of charge-ordered and ferromagnetic phases.
The general aspects of the theoretical description appear to be valid
for any compound with intrinsic phase competition. In high temperature
superconductors, recent experiments have shown the existence of intrinsic
inhomogeneities in many materials, revealing a phenomenology quite
similar to that of manganese oxides. Here, the results for manganites
are briefly reviewed with emphasis on the general aspects. In addition,
theoretical speculations are formulated in the context of Cu-oxides by
mere analogy with manganites. This includes a tentative explanation of
the spin-glass regime as a mixture of antiferromagnetic and
superconducting islands, the rationalization of the pseudogap temperature
$T^*$ as a Griffiths temperature where clusters
start forming upon cooling, the prediction of ``colossal'' effects in
cuprates, and the observation that quenched disorder may be far more
relevant in Cu-oxides than previously anticipated.
\end{abstract}
\end{frontmatter}

$1$: dagotto@magnet.fsu.edu\\
$2$: jburgy@magnet.fsu.edu\\
$3$: adriana@magnet.fsu.edu\\


\section{INTRODUCTION}

The physics of transition-metal-oxides \cite{ref0p0} and other related
compounds appears dominated by states that are microscopically and
intrinsically inhomogeneous in the most interesting ranges of temperatures
and carrier densities. 
The two most relevant
examples are the manganites in the regime of colossal magnetoresistance (CMR),
and cuprates at hole densities in the underdoped region. In manganites
the inhomogeneities arise from phase competition between ferromagnetic
metallic and charge-ordered insulating phases.
In cuprates the competition occurs 
between antiferromagnetic insulating and superconducting or metallic
states. Microscopic
theoretical approaches must consider this phenomenon for a proper
description of manganites and cuprates. Homogeneous states can at best
describe these compounds on large length scales, at a
phenomenological level. 

The experimental evidence
for the presence of inhomogeneous states, particularly in manganites but
also to some extent in cuprates, is simply overwhelming,
and it will not be comprehensibly reviewed here.
Several theoretical studies have also produced considerable evidence
for the intrinsic tendency of electrons in these materials to induce 
competing states that are expected to separate microscopically. 
Some of these theoretical
studies were reported even before the inhomogeneous states were
clearly identified in experiments, highlighting the remarkable
cross-fertilization between theory and experiments that exists in this area of
investigation \cite{ref0,ref0p5}. 
Reviews on this topic, in the manganite context,
are already published \cite{ref1,ref2}.
Moreover, a book on the subject
by one of the authors will be available soon \cite{ref3}. The reader can
find in these references hundreds of citations
covering both theoretical and experimental
aspects. In this short contribution, frequent references to
\cite{ref1,ref2,ref3}
will be made to save space, and reduce the overlap with that 
previous literature. In the cuprate
context, the issue of inhomogeneities is not as universally accepted
as in manganites, although important
recent observations point in that direction.
The potential relevance of electronic phase separation
was remarked in that context many years ago by Kivelson and
others \cite{ref0p5}.
The conference proceedings contained in this volume 
provide one of the best sources of information and references for 
Cu-oxides inhomogeneities. Uemura \cite{ref3p1} has also discussed
extensively
the importance of these inhomogeneities to understand the cuprates.
Readers are encouraged to 
consult the above mentioned literature
to find the relevant papers in this context,
since in the present manuscript we will not
address the details of the remarkable evidence on self-organization
in transition oxides. Only a small subset
of references, mainly by the authors for simplicity, is cited but
the current effort is vast, involving dozens of groups.

This manuscript is divided in two parts. First, we focus on the 
recent proposal \cite{ref1,ref2,ref3,ref4} that manganite
phase competition is in fact the origin of the famous CMR effect,
showing that the inhomogeneities can lead to observable
consequences. In such an analysis,
it will become clear that analogous interesting phenomena can potentially
be observed when any pair of phases are in close competition \cite{ref4}. 
A second kind of phase separation near first-order transitions
is described below, involving phases with the same electronic density
as first observed experimentally by Cheong and collaborators \cite{ref5p1}
and then also theoretically by Moreo et al. \cite{ref18}.
As a natural consequence,
the second portion of the contribution focuses on qualitative predictions
that are made for the cuprates, based on the lessons learned in the
manganite context. In fact, one of the main messages to the readers will be
that manganites and cuprates 
share a similar phenomenology that leads to the potentially
important speculations presented here.


\section{INHOMOGENEITIES IN MANGANITES AND OTHER COMPOUNDS}

Manganites are interesting materials for at least three reasons:


\begin{figure}[ht]
\centerline{\includegraphics[width=.6\textwidth]
{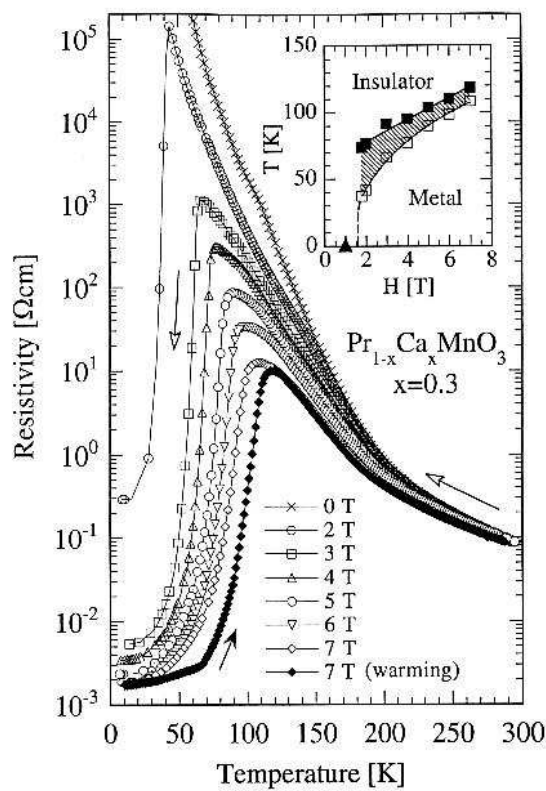}}
\caption[]{
Temperature dependence of the resistivity of $\PCMO$ at $x$=0.3 and 
various magnetic fields. The inset is the phase diagram 
in the temperature-magnetic field plane, with the hatched region denoting
hysteresis. From \cite{tomioka}.
\label{fig1}
}
\end{figure}

{\bf (1)} First,
they have remarkable magnetotransport properties. Figure 1 illustrates
this phenomenon with the example of $\PCMO$ at hole doping 
$x$=0.30\cite{tomioka}. In the absence
of magnetic fields, this material is insulating. However, relatively
small fields of a few Teslas are sufficient to induce a metal insulator
transition at low temperatures. In this regime the resistivity changes
by several orders of magnitude, producing a truly colossal effect. The
state induced by the magnetic fields is a poor metal --the residual zero 
temperature resistivity is high-- and it is ferromagnetic.
Note that fields of order 1T are small when compared with other typical
electronic scales in a single crystal, but unfortunately they are
still too large for applications in computers. 


\begin{figure}[ht]
\centerline{\includegraphics[width=.8\textwidth]
{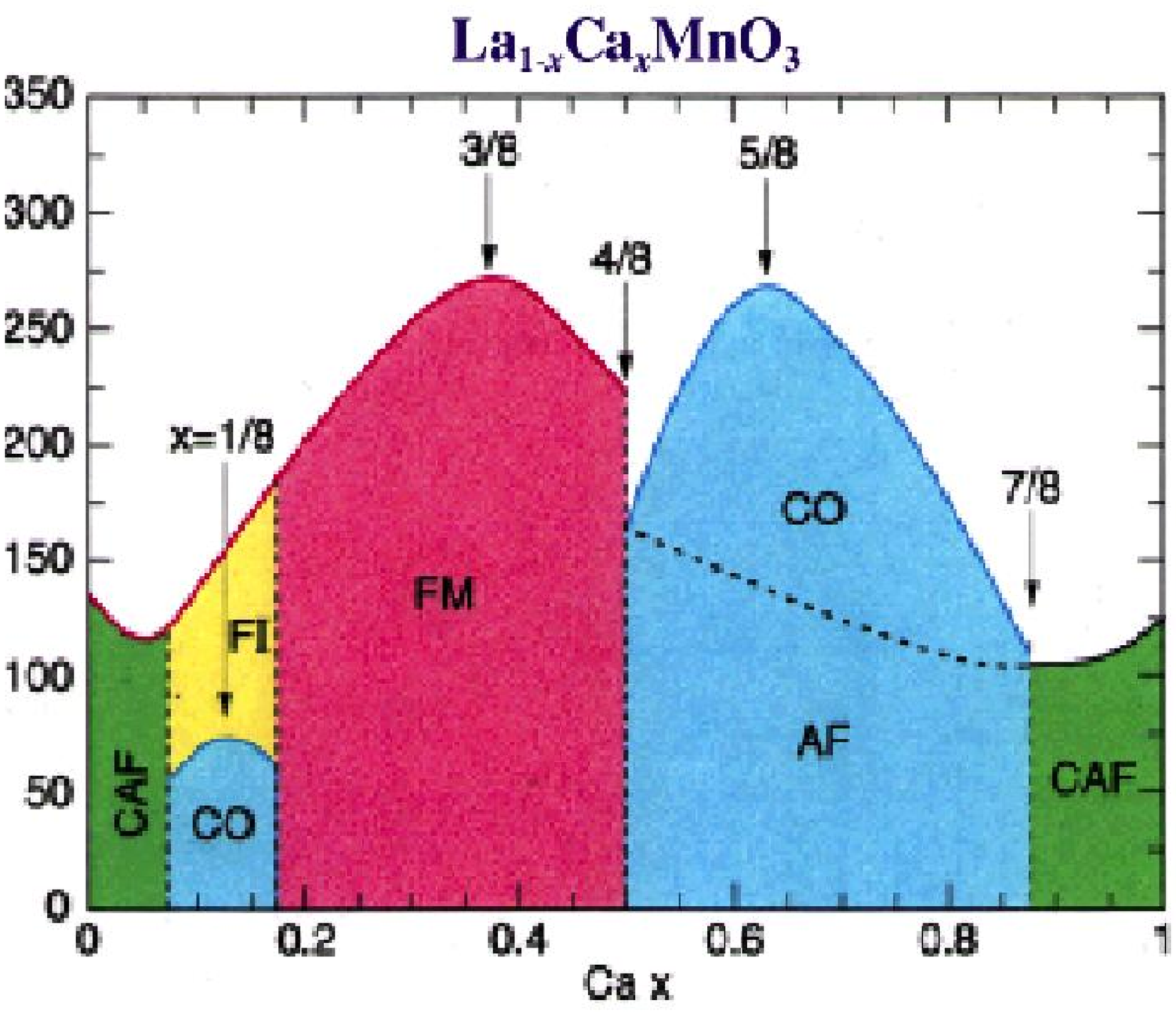}}
\caption[]{
Phase diagram of $\LCMO$, from 
Cheong and Hwang \cite{cheong_hwang}.
The notation is standard. The marked hole density fractions appear to
have more importance than others. 
According to modern theories, the canted regimes (CAF) may
correspond to mixtures of AF and FM regions.
\label{fig2}
}
\end{figure}

{\bf (2)} 
A second reason for studying these compounds is the rich phase diagram
they have. In Fig.~2 the example of $\LCMO$ is shown\cite{cheong_hwang}.
The phase diagram contains a ferromagnetic (FM) metallic phase similar
to the phase induced in Fig.~1
upon the application of a magnetic field. In addition, Fig.~2 exhibits
many other phases, notably a charge-ordered (CO) 
and antiferromagnetic (AF) phase at hole densities $x$=0.50 and beyond.
This phase is also orbitally ordered \cite{ref4p5}, showing here that a new
degree of freedom adds to the charge and spin, leading to complex patterns of
symmetry breaking. It will be argued below
that the competition of these two phases is crucial to understand the CMR
\cite{ref1,ref2,ref3}. The insulating state above the Curie 
temperature $\TC$ is also
nontrivial, and plays a key role in the CMR phenomenon.


\begin{figure}[ht]
\centerline{\includegraphics[width=.8\textwidth]
{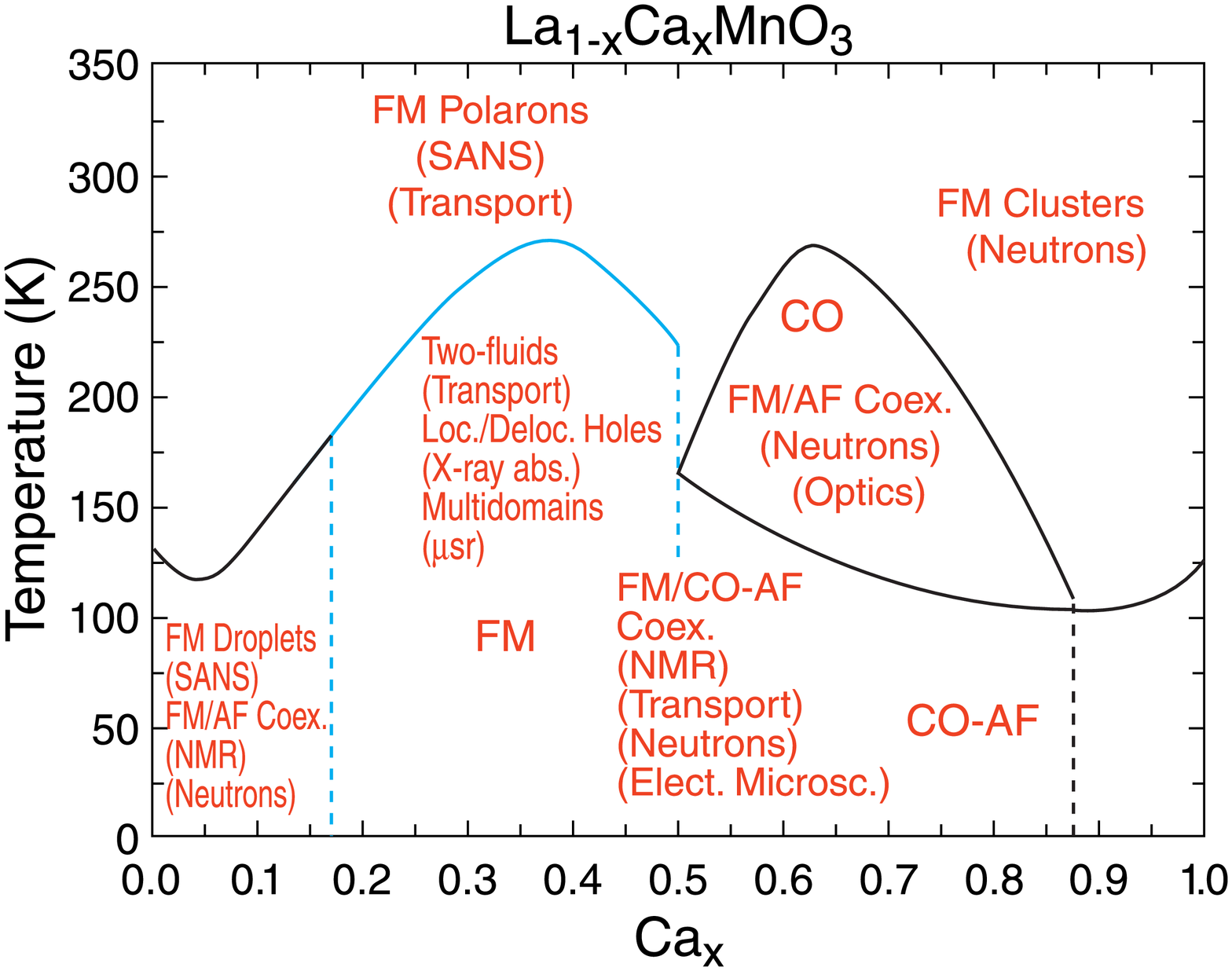}}
\caption[]{
Schematic phase diagram of $\LCMO$, from Moreo {\it et al.} 
\cite{ref1}. The words added refer to statements reproduced from the
experimental literature, at the density and temperature of those
experiments, as well as the technique used (NMR, Neutrons, etc). The use
of words such as droplets, domains, polarons, clusters, and others
indicate a tendency toward inhomogeneous behavior, present in most of
the phase diagram.
\label{fig3}
}
\end{figure}


\begin{figure}[ht]
\centerline{\includegraphics[width=.35\textwidth]
{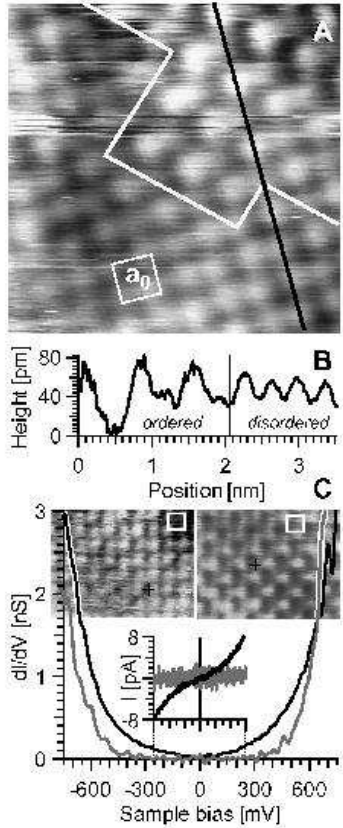}}
\caption[]{
{\bf A.} Atomic scale image of phase segregation at 299~K. From
Renner {\it et al.} \cite{ref5}. The white broken line separates an
insulating charge-ordering region from a more metallic and homogeneous
phase. The compound is $\rm Bi_{1-{\it x}} Ca_{\it x} Mn O_3$ with $x$=0.76. 
{\bf B.} Intensity profile extracted along the straight line shown in A.
Note the difference in amplitude over the Mn sites in the ordered region
due to charge ordering. 
{\bf C.} Spectroscopic signature of phase separation. The spectra were
acquired in the location of the crosses. Two clearly distinct results
are obtained.
\label{fig4}
}
\end{figure}


\begin{figure}[ht]
\centerline{\includegraphics[width=.4\textwidth]
{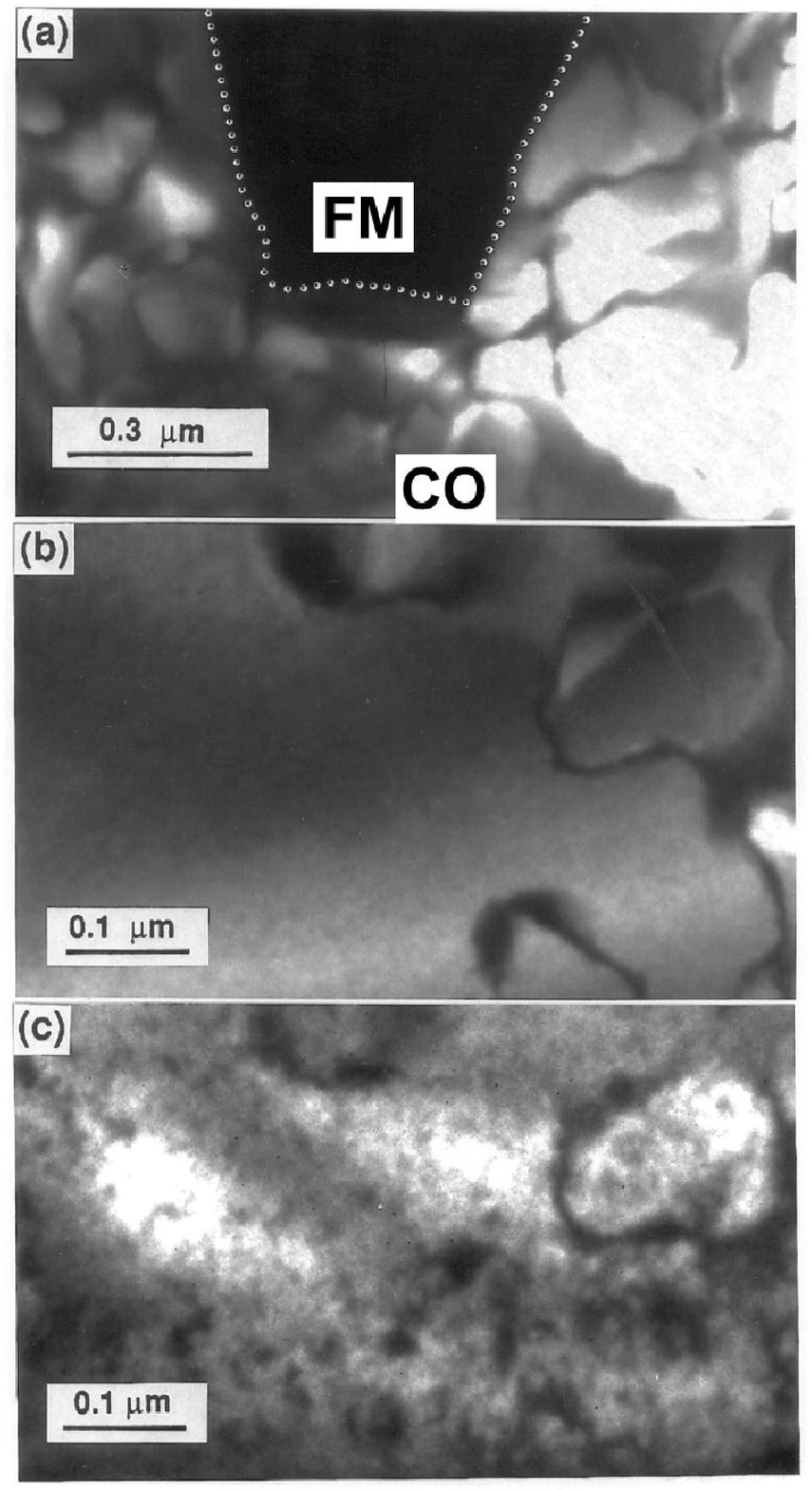}}
\caption[]{
Dark-field images 
for $\LPCMO$, from Uehara {\it et al.}
\cite{ref5p1}. Panel (a) shows the coexistence of insulator CO and metal
FM regions, at 20~K and $y$=0.375. 
Panel (b) corresponds to $y$=0.4 and $T$=17~K, while
panel (c) is also at $y$=0.4 but for $T$=120~K. The latter shows the
development of nanoscale charge disordered domains at $T$$>$$T_{\rm
C}$. 
\label{fig5}
}
\end{figure}

{\bf (3)} 
The third reason for studying manganites is the presence of intrinsic
inhomogeneities, even in the best crystals available. Figure 3 is 
reproduced from \cite{ref1}, where the phase diagram of $\LCMO$ is shown
again, but this time including brief 
statements taken from the experimental literature
that highlights the presence of inhomogeneities. Words such as ``polarons'',
``clusters'', ``multidomains'', ``AF-FM coexistence'' and related ones
are ubiquitous in the experimental literature. These materials appear to
have a tendency toward an inhomogeneous state, and 
typical length scales mentioned often
are of a few lattice spacings (nanoscale). The pioneering 
experiments in this context are reviewed in Refs.~\cite{ref2,ref3}, and the
details will not be repeated here. Among the most recent
experiments are those of Renner {\it et al.} \cite{ref5} using 
Scanning Tunneling Microscopy (STM) applied to $\BCMO$ in the regime of
high hole doping and room temperature. The results are reproduced in
Fig.~4. They show atomic resolution features in the surface of this
compound and the presence of two types of charge ordering (upper panel):
checkerboard and homogeneous. The former is associated with the CO/AF
state, while the latter is presumably the FM state. In Fig.~5, dark-field
electron microscopy results for $\LPCMO$ are also reproduced, this time
from Uehara {\it et al.} \cite{ref5p1}. This material has a clear
competition CO/AF vs. FM that can be tuned by varying the relative amount
of La and Pr. The upper panel illustrates the low temperature results
showing the presence of coexisting metallic and insulating regions.
Their size is very large, at the submicrometer scale. If this type of
experiment is indeed testing intrinsic properties of crystals, this
large length scale introduces limitations on the theoretical considerations.

It will be argued later 
that motivations (1), (2), and (3) are actually \emph{interconnected}.
The rich phase diagram (2) causes phase competition and concomitant
inhomogeneities (3), which themselves induce CMR (1). The details are in the
following sections.


Recent investigations have shown that inhomogeneities are present
in other compounds as well. Consider, for instance, the widely publicized
remarkable STM results by the group of Davis \cite{ref5p5}, where a wide
distribution of $d$-wave superconducting gaps was observed at the surface
of Bi2212 in the superconducting regime. The size of the clusters was
found to be in the nanometer
range. This occurs both in the optimal and
underdoped regimes. Note that the universality of this property is still under
discussion and inhomogeneities may not be as prominent 
in other cuprates such as  YBCO \cite{ref5p7}. But at the very least,
measurements in the much studied Bi2212 material must be reanalyzed in view of
the recently discovered inhomogeneities. In addition, there are 
dozens of papers that have
reported stripe-like structures, particularly in $\LSCO$ in the
underdoped regime. This is another manifestation of the
intrinsic tendency toward microscopically inhomogeneous states (the list
of references simply too long to be reproduced here). Finally, recent
scanning SQUID microscopy 
study by Iguchi {\it et al.} \cite{ref6} have reported the existence of 
diamagnetic activity above $\Tc$ in $\LSCO$. The phenomenon
is found at temperatures as high as 80~K, which is remarkable!
The size of the superconducting islands was found to be as large as
several micrometers.

Adding to the evidence of inhomogeneous states 
in cuprates, recently indications of clustered states in
Eu-based compounds have been reported in \cite{ref6p1}. In this reference, the
analogies with manganites were discussed and emphasized. 
Previously, it was widely
believed that ferromagnetic polarons in Eu semiconductors were 
responsible for their properties. However, the
most recent results \cite{ref6p1} suggest a distribution of clusters of
different sizes that may contain several carriers each. These are
\emph{not} polarons (one carrier
with a spin distortion around) but more complex structures. The
same observation regarding the relevance (or lack of it) of polarons
is valid for cuprates and manganites as well. Their inhomogeneities
cannot be visualized as ``polarons''.

Finally, in ruthenates the possibility of orbital ordering has been
recently proposed \cite{ref7}. In addition, large MR effects have been
unveiled in bilayer Ru-oxides \cite{ref7p1}. 
Phase competition was also found in the single layer ruthenate \cite{ref7p2}.
It seems that (Ca,Sr)-based ruthenates may behave similarly as other
materials mentioned here. Nickelates, cobaltites, and other compounds
add to the list where the inhomogeneities dominate in their ground state
properties (for a far longer list than presented here, 
and more details, see Chs. 20 and 22 of \cite{ref3}).


\section{ELECTRONIC PHASE SEPARATION IN MANGANITES}

On the theory front, most of the work carried out in the area of
manganites uses models with two relevant degrees of freedom. One of
them are the localized $\t2g$ spins and the other are the
mobile $\eg$ carriers. The latter has two relevant orbitals
per Mn-ion. As a consequence, typical Hamiltonians involve: (1) The hopping
of $\eg$ electrons regulated by a 2$\times$2 
hopping matrix with an overall scale $t$ (the hopping amplitude). (2) A local
Hund coupling $\JH$ that enforces alignment of spins between the $\eg$
and $\t2g$ spins (experimentally the spin of $Mn^{3+}$ is known to take the
maximum value $S$=2, compatible with a large Hund coupling). 
(3) A relatively small Heisenberg antiferromagnetic
coupling $\JAF$ between the localized $\t2g$ spins, which, however, plays an
important role in regimes where there are competing states of similar energy.
In addition, either on-site Coulombic interactions with couplings
$U$, $U'$, and $J$ in the standard notation (Ch. 4 of \cite{ref3}), 
or an interaction between
the electrons and Jahn-Teller (JT) phonons --regulated by a dimensionless
coupling $\lambda$-- are incorporated. In the approach favored by the
present authors, the latter is used. In addition, 
if one further assumes that
the phonons are classical, Monte Carlo simulations can be carried out
without substantial technical difficulties \cite{ref8}. Moreover, evidence
shows that Coulomb interactions can be mimicked by combined large
$\lambda$ and $\JH$ (Ch.8 of \cite{ref3}), 
and it is expected that Coulombic or JT dominated
models will lead to analogous behavior, at least at low temperatures.

Using this type of models, Yunoki {\it et al.} \cite{ref0,ref8} and
later several others (see \cite{ref2,ref3} for references)
have unveiled a clear signal of ``electronic phase separation'' in
the manganite context. 
This type of phase separation manifests as
a discontinuity in the density of carriers as the chemical potential $\mu$
varies, for example in a Monte Carlo simulation. It was observed that even
if $\mu$ 
changes smoothly, there are electronic densities that can never be stabilized.
If the system is forced to have such densities (for example working in
the canonical ensemble), then the ground state separates spontaneously
into two macroscopic regions, each carrying the phase at the extremes
of the density discontinuity found varying $\mu$. At least for the one-orbital
model (restricting the number of relevant $\eg$ orbitals to one), the
separation involves AF hole-undoped and FM hole-doped phases \cite{ref0}. For
two orbitals the situation is more complicated and it involves the orbital
degree of freedom \cite{ref8}. Electronic phase separation is a robust
effect, and its presence in manganite models is not in doubt. The reader can
find a discussion and the actual evidence of this behavior in
\cite{ref1,ref2} and Ch. 6 of \cite{ref3}. The
word electronic in front of phase separation is used here to remark the
different densities of the two competing phases. It is widely accepted
that if the tail $1/r$ of the Coulomb interaction is incorporated, the
macroscopic phase separation mutates into a microscopic effect, with the
formation of clusters in the \emph{nanoscale} range
(see \cite{ref0p5} and Ch. 6 of \cite{ref3}). This has implications for
the rationalization
of the results of Uehara {\it et al.} \cite{ref5p1} where submicrometer
clusters were reported.

It is important to state that electronic phase separation in models
of cuprates has been proposed even before the recent work in manganites.
In the cuprate context,
the proposed separation involves the AF insulator and either a metal or
a superconductor (see for instance 
the work of Kivelson and collaborators in Ref.\cite{ref0p5}). Phase separation 
appears clearly in the famous $t$-$J$ model, 
although it is still under much discussion whether it occurs
at realistic values of $J/t$ \cite{added}.


\begin{figure}[ht]
\centerline{\includegraphics[width=.48\textwidth,angle=270]
{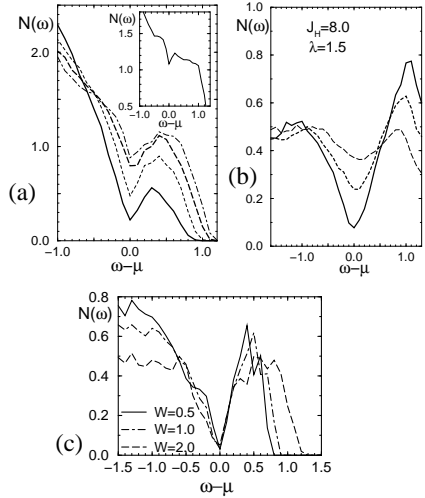}}
\caption[]{
(a) DOS of the one-orbital model using a 10$\times$10 cluster at
infinite Hund coupling, and temperature $T$=1/30 in units of the hopping.
The four lines from the top correspond to densities of mobile carriers
equal to 0.90, 0.92, 0.94 and 0.97. The inset has results at density 
0.86. (b) DOS for the two-orbital model using a 20-site chain, 
electronic density 0.7, Hund coupling 8, and $\lambda$=1.5. The three
lines at the chemical potential from the top correspond to temperatures
$T$=1/5, 1/10, and 1/20. Both (a) and (b) are reproduced from 
Moreo {\it et al.} \cite{ref9}, where more details can be found.
(c) DOS in the presence of disorder, 
from \cite{ref18}. $W$ is the strength of disorder, and shown are
results for the one-orbital model on a chain of 20 sites, with 
temperature $T$=1/75 (hopping units), Hund coupling 8, and density 0.87.
This corresponds to a regime of phase separation for zero disorder.
The disorder stabilizes the system, and creates a pseudogap.
For more details see \cite{ref18}.
\label{fig6}
}
\end{figure}

An interesting observation is that in regimes of temperatures, couplings,
and densities that have cluster coexistence, there is a
\emph{pseudogap} in the density of states (DOS)  (see Fig.6).
This occurs, for example, at densities where
phase separation exists by lowering the temperature to $T$=0.
Intuitively, at intermediate temperatures precursors of phase separation
must be present in the form of coexisting clusters.
The existence of this pseudogap feature
was remarked theoretically in \cite{ref9}. Photoemission experiments have also
unveiled a similar behavior in bilayer manganites \cite{ref10}. It is expected
that pseudogaps would appear in the density of states in the regimes of
inhomogeneities, as a natural consequence of the competition between
a metal (flat DOS) and an insulator (gapped DOS). As discussed later,
this should occur in CMR manganites above the Curie temperature as well, 
since phase competition is expected in that regime.
A pseudogap is well known to exist in underdoped
copper oxide materials as well.


\section{GENERAL ASPECTS OF PHASE COMPETITION IN THE PRESENCE OF QUENCHED
DISORDER}

The discovery of electronic phase separation in manganite models
described in the previous section, and the resulting nanoscale coexisting
clustered-state upon the introduction of $1/r$ Coulomb effects, provides
a first approximation toward the understanding of the physics of
manganites (Fig. 3).
This possible explanation is robust on theoretical grounds and compatible with
experimental data. However, it is important to analyze phase separation
in more general terms. In fact, the experiments of Uehara {\it et al.}
\cite{ref5p1} show clusters of sizes in the submicrometer scale
involving two phases (varying $y$ in $\LPCMO$ it is possible to interpolate
between FM and CO states at  constant $x$). The electronic density of
these coexisting clusters is likely the same since different densities
would lead to large energy penalizations due to the accumulation of charge.


\begin{figure}[ht]
\centerline{\includegraphics[width=.48\textwidth]
{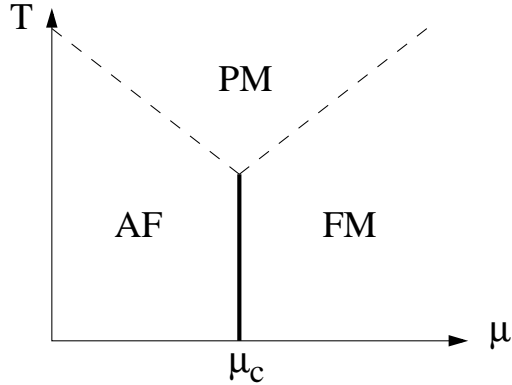}}
\caption[]{
Schematic representation of the electronic phase separation phenomenon
in the $T$-$\mu$ plane. The notation is standard.
\label{fig7}
}
\end{figure}

The results of the previous section can be reformulated as a first order
transition as a function of the chemical potential
$\mu$, as shown in Fig.~7. Let us now consider
a similar first-order transition, but now varying an arbitrary parameter
instead of $\mu$. This transition could occur at constant electronic
density, changing for example the hopping amplitude $t$ (which can 
effectively be done by chemical substitution as in $\LPCMO$). 
In this case the $1/r$ Coulomb interaction will not lead to such dramatic
consequences (nanoclusters) as in the electronic phase separation case.
However, \emph{quenched disorder}
produces interesting results.  To carry out this type of calculations, 
recently Burgy {\it et al.} \cite{ref4} used an Ising 
spin ``toy model'', with couplings $J_1$, $J_2$ and $J_4$
at distances 1, $\sqrt{2}$, and $\sqrt{5}$ lattice spacings. It is expected
that the general aspects of the results will not be severely
affected by the details of the model. The Hamiltonian was selected such
that two phases are in competition in the model under investigation.
In Fig.~8 they are called phases ``$O_1$'' and ``$O_2$'' (in practice
FM and AF  ``collinear'' phases, respectively). The reader should not
be deterred by the fact that the present convention for
``$O_1$'' differs from that in \cite{ref4}. The model is invariant
under the transformation ($J_1$,$J_2$,$J_4$) $\rightarrow$
(-$J_1$,$J_2$,-$J_4$), hence the FM and AF states are interchangeable.
The couplings are selected such
that in the absence of disorder the phase diagram is as shown in Fig.~8
with dashed lines. A first-order transition separates the two phases
at low temperatures.
This is  similar to Fig.~7 but now varying one of the couplings
($J_2$) in the model, rather than the chemical potential. 
The critical value is $J_{\rm 2c}$=0.7$J_1$, but its actual value
is a numerical detail not believed to be of relevance.


\begin{figure}[ht]
\centerline{\includegraphics[width=.75\textwidth]
{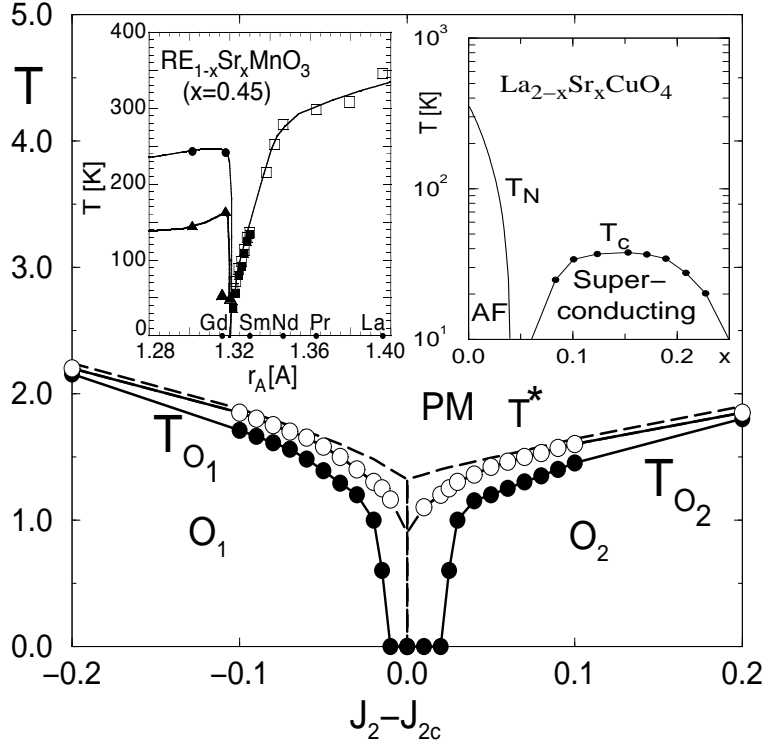}}
\caption[]{
Phase diagram of the ``toy model'' of Ising variables with
$J_1$-$J_2$-$J_4$ \index{model $J_1$-$J_2$-$J_4$}
couplings in two dimensions ($J_4$=0.2$J_1$ to induce a clear first-order
transition at low temperature, and $J_1$=1 provides
the energy scale). This phase diagram is believed to 
represent the behavior of any pair of competing phases.
Disorder is introduced into the
coupling $J_2$ that allows the system to change from phase 1 to phase 2,
which in this case are ferromagnetic and collinear, respectively.
Details of the definition of the disorder, lattices and techniques used,
can be found in the original reference Burgy {\it et al.} \cite{ref4}.
$T_{\rm O_1}$ and $T_{\rm O_2}$ are the true ordering temperatures,
while $T^*$ \index{clean limit $T^*$}
is the clean-limit ordering temperature, which survives
as a rapid ``crossover'' for cluster formation when disorder is introduced.
The insets are phase diagrams of Mn-oxides on the left 
(private communication from Y. Tokura and Y. Tomioka), and the single
layer cuprate on the right.
\label{fig8}
}
\end{figure}

Quenched disorder is introduced by adding a random component to $J_2$
since it is this coupling that must be varied to transform 
from one phase to the other. The random component is taken from a
box distribution centered at zero, of total width $W$.
From the study of Imry and Wortis \cite{wortis} it is to be
expected that disorder will transfrom a first-order transition
into a continuous one. Simulations indicate a rich phase diagram
when the two phases compete and disorder is introduced.
The results for
two typical values of $W$ are in Fig.~8, reproduced from \cite{ref4}.
At values of $J_{2}$ far from the region of competition, the disorder
strength used is not sufficient to alter the value of the critical
temperatures. However, in the region of competition near $J_{\rm 2c}$, 
far more dramatic effects are observed. For ``weak'' disorder, both
critical temperatures are appreciably reduced around $J_{\rm 2c}$, although
still the general shape of the phase diagram resembles bicritical or
tricritical behavior. This is one class of results that may be observed
in experiments, as reported recently in 
$\rm {Pr_{0.55}(Ca_{\it 1-y}Sr_{\it y})_{0.45}Mn O_3}$ \cite{ref11}.
On the other hand, for ``large'' disorder
the reduction in the critical temperatures is far more dramatic, and it
leads to a region \emph{without} long-range order even at $T$=0.
An example of this could be 
$\rm {(La_{\it 1-x}Tb_{\it x})_{2/3}Ca_{1/3}Mn O_3}$, reported in
\cite{teresa}.

If we are correct in assuming that the general aspects of the problem
remain for more realistic competing phases, it is then predicted that
in materials with phase competition two types of phases diagrams are
to be expected for different strengths of the disorder (namely,
the weak and strong disorder 
cases of Fig.~8). In the insets of the same figure, two experimentally
determined phase diagrams are shown with features reminiscent of those in
the theoretical calculation. On the right, is the well known phase
diagram of $\LSCO$ that in the underdoped regime shows a clear depletion
of both the N\'eel and superconducting (SC) critical temperatures, forming a
region widely known as the ``spin glass''. It is tempting to speculate
that this region is actually produced by phase competition between 
AF (perhaps containing stripes) and SC states. The actual form of the state
is discussed below. On the left inset of Fig.~8 is the phase diagram of
the Sr-based $x$=0.45 manganite as determined by Tomioka and Tokura
\cite{ref11}.
Note the presence of a deep reduction in the critical temperatures, generating
a feature that resembles ``quantum critical'' behavior. Both insets have
qualitative similarity with the theoretical study, and it is natural to believe
that this is not accidental. 


\section{STATES OF RELEVANCE IN THE REGION OF PHASE COMPETITION AND
PREDICTION OF $T^*$}

What sort of states are induced in the region of competition upon the
introduction of disorder? Typical results are shown in Fig.~9, where
averages over a few hundred Monte Carlo sweeps are shown. On the left,
the state is dominated by the white color. In our convention that means
that in most of the sites of the lattice the order parameter is not well
developed, namely, as time evolves the values of the
order parameters fluctuate leading to a net zero result. This is 
the picture of a standard paramagnet, and it occurs at temperatures above
the original critical temperature of the ``clean'' (not disordered) limit.
{\it This temperature will play a key role in the
following and we denote it by $T^*$}. This is a Griffiths temperature.
Griffiths effects appear to be substantially magnified when disorder is
introduced in regions of phase competition \cite{ref4,salamon}.


\begin{figure}[ht]
\centerline{\includegraphics[width=.8\textwidth]
{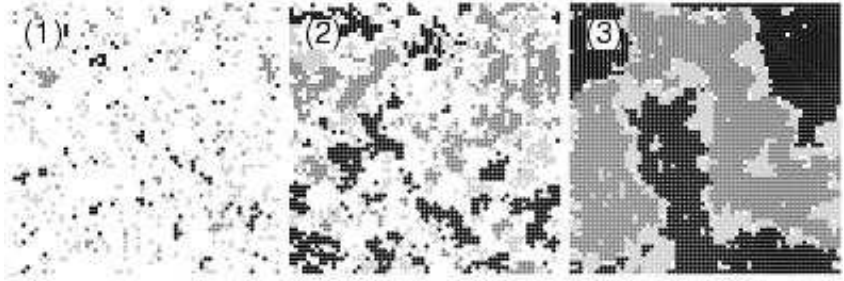}}
\caption[]{Typical spin configurations representative of dominant states in
the  toy model, generated by the MC simulation. Shown are
averages over nearly 100 MC sweeps. The coupling and temperature
for the three cases (1-2-3) are those marked in Fig.\ref{fig8}.
The conventions used are the following: the darkest regions
correspond to the FM phase with positive order parameter; the next dark
tone is FM with negative order parameter; the light grey is the competing
collinear phase; while white corresponds to a paramagnetic region.
The original colors can be found in Burgy {\it et al.} \cite{ref4}.
\label{fig9}
}
\end{figure}

In the middle of Fig.~9 the state between the actual critical
temperature $T_{\rm O_1}$ and $T^*$ is presented. This state will lead
us to a rationalization of the CMR effect, as described below. It still
contains some paramagnetic (white) areas, but now it is clear that 
there are regions that have quasi-static \emph{local order} and
clusters are formed.
These clusters are denoted in the figure with three grey colors as indicated
in the caption. Some of them have the order that will become truly
dominant below $T_{\rm O_1}$ (which is actually a Curie temperature in
the example considered). However, these clusters can randomly have
a positive or negative order parameter, leading to a
\emph{globally disordered}
state. If Heisenberg variables --instead of Ising-- would have been 
considered, then the orientation of the local order parameters of the
clusters would point in arbitrary directions, but still leading to a global
cancellation. In addition to these  clusters, in the figure
one can observe small islands of the competing phase, namely, the phase
that will become stable upon further increasing couplings. The 
fluctuations in the disorder create regions where $J_2$ is locally
larger than the critical value, and the other phase is stabilized.

It is easy to imagine that transport in such a complicated environment
is quite complicated. Consider, for instance, a spin-up electron 
that crosses the sample coupled to the ``toy model'' states of
relevance by a Hund coupling, as it occurs when $\eg$ and $\t2g$ degrees
of freedom are considered. Movement within the spin-up locally ordered
regions should be nearly ballistic for small clusters. However, for
the spin-down regions as well as the competing phase regions (that contain
spins up and down in stripes) the movement of the spin-up electron
is not favored. They act as ``insulators'' for the up species of electrons.
These insulating regions will increase the resistivity dramatically. There
is no obvious easy channel for the flow of charge from one side of the
sample to the other in this context. We will see below that this translates
into a huge resistivity.

Finally, as the temperature is lowered further, then either the spin up or
down clusters dominate, a percolation occurs and the dominance of
one ``color'' is found in simulations 
(as shown in Fig.~9-3). This state will be favorable for
transport of at least one spin species of electrons, although it is still
inhomogeneous. Note the interesting formation of domain-walls in the FM
order parameter, through the stabilization of the competing phase \cite{ref4}.


\section{GENERALIZATION OF TOY MODEL RESULTS TO REAL MANGANITES}

Calculations such as those reported in the previous section cannot be
performed directly using realistic models for manganites. These models
involve many degrees of freedom, some of them quantum mechanical
(e.g. the $\eg$ electrons).
Such complex system cannot be easily simulated in clusters large enough
to see percolative physics (although they can be studied fairly well
to obtain phase diagrams in the absence of disorder). 
However, the results obtained with the
toy model were easy to understand and they seem general enough to be
valid under several other circumstances as well. Thus, we believe that
in real manganites the competition between the FM and CO/AF states in
the presence of sources of disorder
also leads to a phase diagram as shown in Fig.~8, and to states as in Fig.~9
simply changing the labeling of the phases. As a consequence, 
two interesting conjectures can be made for manganites inspired by the
study of simpler systems: (1) We believe that the CMR state has
a ``clustered'' structure, with preformed FM and CO/AF clusters and
even PM regions. Figure 10 shows a sketch of this state. It
has no global net moment, but locally there is order (see also \cite{ref5p1}). 
(2) There has to be a $T^*$ scale in Mn-oxide real systems that
correspond to a Griffiths temperature where clusters start forming.
This temperature is larger than the true ordering temperatures.


\begin{figure}[ht]
\centerline{\includegraphics[width=.6\textwidth]
{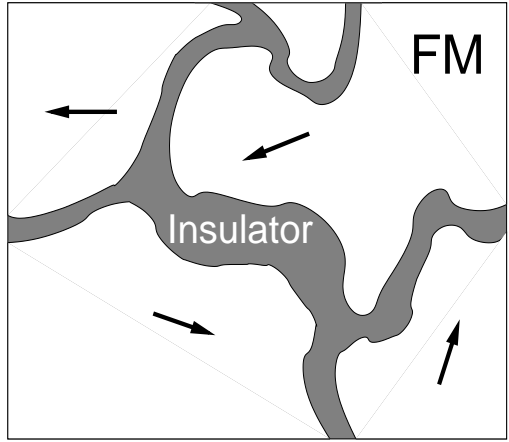}}
\caption[]{
Proposed state for manganites in the CMR regime. FM
clusters are locally formed, but with random orientations of the
order parameter. The insulator forms walls between the FM metallic
regions.
\label{fig10}
}
\end{figure}

To test these assumptions, calculations must be carried out to check
the presence of a CMR effect in a clustered state. This will be the
goal of the next section. In addition, the $T^*$ new scale should
be observable in real materials. Some experimental results are
described below that have reported results compatible with the clustered
state and the existence of $T^*$.


\section{POSSIBLE EXPLANATION OF THE CMR EFFECT}



\begin{figure}[b]
\centerline{\includegraphics[width=.85\textwidth]
{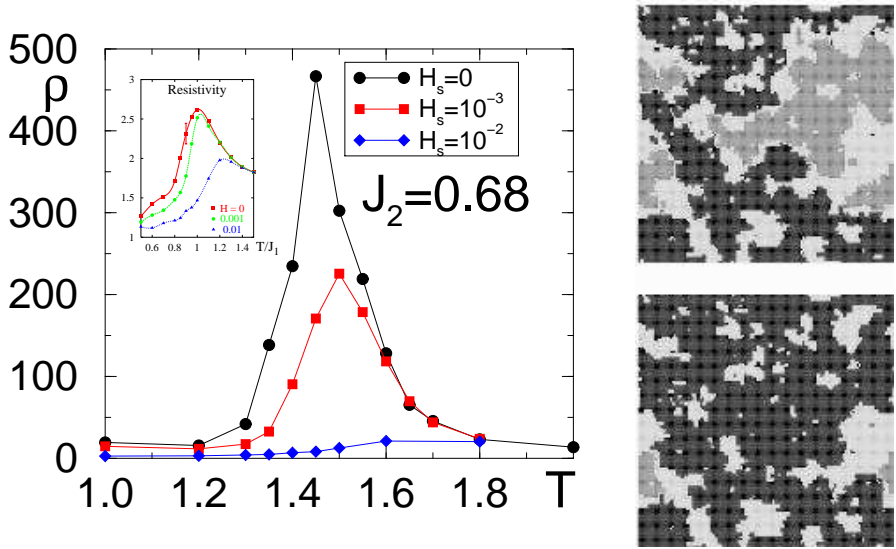}}
\caption[]{
Resistance close to $J_{2c}$, obtained from the MC 
configurations of the toy model $J_1$-$J_2$-$J_4$ by a suitable
translation to manganite language of the many phases,
supplemented by mild assumptions on the individual
resistances that form the resistor network (see text and Ref. \cite{ref4}). 
Note the large value of the zero-field resistance in the intermediate
temperature region with preformed clusters, where there is a global
cancellation of the order parameters (black is FM with positive moment,
dark grey is FM with negative moment and light grey is the phase competing
with FM). Resistances including an external
field $H_s$ (in units of $J_1$) are also shown. There is a
strong dependence of the resistance with external field, leading
to a huge MR ratio, comparable to experiments. The snapshots on
the right correspond to typical configurations before (upper panel)
and after (lower panel) a field $H_s = 10^{-2}J_1$ is turned on
(for details see \cite{ref4}). The inset shows similar behavior in 3D
(see Fig. 12).
\label{fig12}
}
\end{figure}

From the spin toy model discussed before, it is not possible to obtain
the resistivity directly. However, it is possible to make reasonable 
assumptions that would allow us to obtain a rough estimation of that
resistivity. As described in the previous section, we will consider
electrons with either spin up or down moving in the background of the
spins generated by the Monte Carlo simulations (represented in the
various regimes by the three typical states shown in Fig.~8). For each
state, simple rules can be established that would allow us to write
a {\it random resistor network} approximation to the problem.
The reader
can find some details in \cite{ref4}, but the idea is simple: from the
perspective of, say, the spin-up electron then (1) a low resistance in 
the effective network should
connect FM regions with the positive magnetization; (2) a high (or
infinite) resistance links the positive and negative magnetization FM
regions (due to the Hund coupling spin-up electrons do not propagate
in the negative moment FM areas); and (3) an intermediate resistance
links the FM positive moment and PM regions. Analogous rules can be
setup for spin-down electrons. By this procedure, for each Monte Carlo
generated ``snapshot'', a resistor network is constructed, and then
solved iteratively using the Kirchoff equations.

The resistance (or resistivity) vs. temperature obtained by the procedure
outlined  in the previous paragraph is shown in Fig.~11, for
a coupling $J_2$ close to $J_{\rm 2c}$. In the absence of magnetic
fields, a \emph{huge} peak is found at intermediate temperatures in 
qualitative agreement with experimental results, and also with the
theoretical expectations as described in the previous section. The
state between $T_{\rm O_1}$ and $T^*$, with its clustered and messy
structure, is very detrimental for transport of charge, and the
results of the resistor network approximation confirm this guess.


The most spectacular result is the dependence of the resistance with
magnetic fields, also contained in Fig.~11.  Consider, for example,
a tiny field of just $10^{-3}$$J_1$ in the natural units of the problem.
This field is found to reduce the peak in the resistivity by 50\%,
an enormous effect! For a field of $10^{-2}$$J_1$ the
peak has basically disappeared in the scale of Fig.~11. This effect
found in simulations is quite similar in scale to those reported
experimentally. Similar results are found in 3D simulations (Fig. 12),
although of smaller magnitude (work is still in progress in this
contex, see \cite{burgy3d}).

\begin{figure}[hb]
\centering
\rotatebox{90}{
\includegraphics[width=.45\textwidth]
{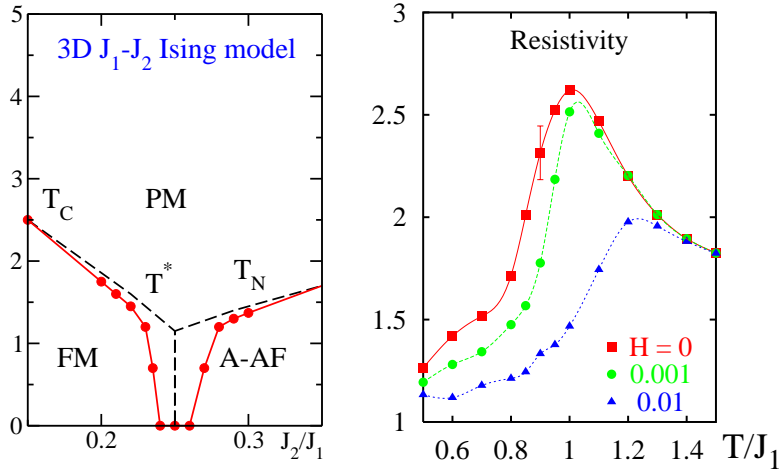}}
\caption[]{
{\it (left)} Phase diagram of the $J_1$-$J_2$ toy model using a
three dimensional lattice. The couplings were selected
such that ferromagnetic and collinear (or A-AF) phases are in
competition. The $T^*$ scale is indicated. Results are quite similar to 
those in two dimensions.
{\it (right)} Resistivity vs temperature, at the external fields indicated.
A large magnetoresistance is observed. The calculation was done using
similar resistor network rules as in two dimensions. Details, such as
couplings used and strength of disorder, can be
found in \cite{burgy3d}.
\label{3d.1}
}
\end{figure}

What is the physics behind the results in Fig.~11? Consider also in Fig.~11
(right panels)
a typical state found in the simulations at zero field, in the
temperature range of interest. As described before, it contains
regions with positive and negative magnetization as well as insulating
domains.  The key issue for the present discussion is the relatively robust
size of the preformed ferromagnetic
clusters. Suppose the magnetic field has a sign such that it
favors the positive magnetization. The negative magnetization clusters
involve dozens of spins and in this respect they
behave like giant effective spins. Such a giant spin can rotate in spin space
under the influence of a tiny magnetic field. The effect is large not
because the field is large, but because the preformed effective spin
is large! In fact the bottom right panel of Fig.~11 shows the net effect
of adding a magnetic field of $10^{-2}$$J_1$. The regions with
negative magnetization have flipped to positive, the insulating regions 
separating positive from negative magnetizations have melted away
(since they are no longer needed), and now several channels for transport
are opened for spin-up electrons. A small magnetic field produces
huge changes in transport due to the existence of preformed ferromagnetic
clusters! We believe that these ideas have captured the essence of the 
CMR paradox, although much more work is certainly needed to fully refine them.


\section{PHASE COEXISTENCE ABOVE $\TC$ and EXPERIMENTAL DETERMINATION OF $T^*$}

At present, several exciting experiments are being carried out to
study the existence of a $T^*$ new scale in manganites. We cannot review
them all here for lack of space, but the reader can consult Ch. 19 of
\cite{ref3} for a more detailed discussion, preliminary information, and 
additional citations. 
For our purposes, here we will only describe the recent work of
Adams {\it et al.} \cite{ref12} where neutron scattering results for $\LCMO$
with $x$=0.30 were reported in a wide range of temperatures. Some
results are in Fig.~13. The authors of Ref.\cite{ref12} studied diffuse
scattering in the vicinity of a Bragg peak. Some features at
particular values of the momenta are identified
as caused by the presence of ``uncorrelated'' 
Jahn-Teller polarons. The intensity of one of those momenta vs. temperature
is in Fig.~13 (a). 
Below the Curie temperature the signal is small, while
in the range experimentally
investigated the intensity is nearly \emph{constant} above $\TC$.
This behavior does not follow the resistivity of the system, which 
has a large
peak at $\TC$, and it rapidly decreases both below and above that
ordering temperature. Then, the uncorrelated polarons (namely,
a state dominated by a gas of independent fairly heavy polarons) is 
{\it not the state of relevance for manganites,
since it does not correlate with the resistivity}. 


\begin{figure}[ht]
\centerline{\includegraphics[width=.85\textwidth]
{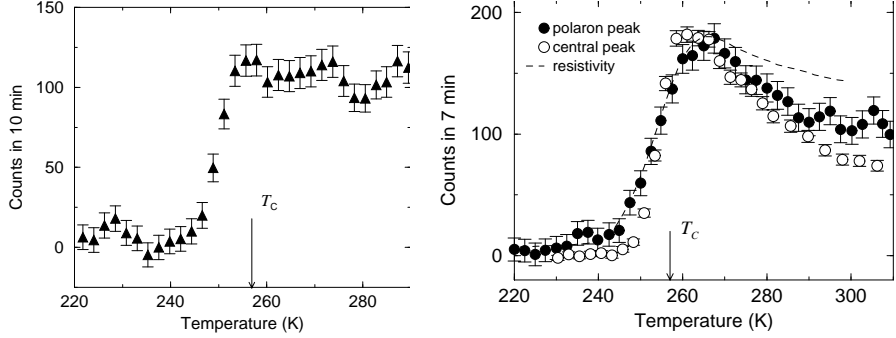}}
\caption[]{
(a) Temperature dependence of the diffuse polaron scattering (uncorrelated
polarons) for $\LCMO$ ($x$=0.30). Note that above $T_{\rm C}$
the results are nearly temperature independent in the range investigated. 
(b) Temperature dependence of the intensity of the ``polaron peak''
--corresponding to charge ordering-- compared to the central peak 
intensity discovered by Lynn {\it et al.} \cite{lynn}, as well as the
resistivity. The data have been scaled so the peak heights match.
It was concluded that ``The similarity of the data indicates a
common origin''. From Adams {\it et al.} \cite{ref12},
where more details can be found.
\label{fig13}
}
\end{figure}

On the other hand, in Fig.~13 (b), results of the same experiment
but at other momenta are shown. They correspond to the signal attributed
to ``correlated polarons'', which manifest as a weak peak in the vicinity
of the dominant Bragg peak. This feature in the neutron intensity indicates
that polarons are not independent, but they form a structure that seems
to resemble closely the CE-state of half-doped manganites \cite{ref12p5}.
Then, the correlated polaron signal should be more properly referred to as
CE-clusters or charge-ordered clusters, and they correspond to
small islands of a phase (CO/AF) that becomes stable by changing the
chemical composition. The key result of Ref.\cite{ref12}
(Fig.~13 (b)) is that the intensity corresponding to the
charge-ordered clusters behaves as a function
of temperature quite similarly as the resistivity does. 
Adams {\it et al.} \cite{ref12}
write that {\it ``The similarity of the data
indicate a common origin.''} Then, the coexistence
of CE-like clusters, with the FM clusters known to exist in the same
regime \cite{ref13} and, probably, paramagnetic regions as well, forms
a complex state whose existence is correlated with the anomalous behavior
found in transport measurements. 
Following the neutron peak intensity related with charge-ordered clusters
vs. temperature would allow for a determination of $T^*$ when the signal
vanishes.
These conclusions are supported by a variety of
measurements by several groups in addition to 
those described here (for a list see Ch. 19 of \cite{ref3}). $T^*$
in materials such as $\LCMO$ ($x$=0.30) appears to be located in the
neighborhood of 400~K \cite{ref13}. Adding to these results, recent
studies by Argyriou {\it et al.} \cite{ref14} have also reported
a $T^*$ in bilayered manganites that appears to correspond to a glassy
transition, at a temperature well above the ordering temperatures. 
Work in this important and exciting subarea of manganite physics is just
starting, and many surprises will likely be found in the near future.


\section{LESSONS FOR THE CUPRATES}

The results for manganites described above, with inhomogeneities clearly
found both in experimental and theoretical investigations, appear to 
originate in phase competition. The general considerations mentioned during
the discussion of this problem indicate that phase competition between
any pair of fairly different ordered phases should lead to a similar
phenomenology. In particular, the results of Fig.~8 should apply to
the superconductivity vs. AF insulator competition in high temperature
superconductors. As a consequence, by mere analogy with manganites it is
possible to list some properties that cuprates may have if indeed they
behave similarly as other transition metal oxides \cite{ref4}. The results
of the present discussion when applied to cuprates can
be labeled as ``speculations'' at this point, since it is difficult to
carry out detailed calculations for Cu-oxides. However, they are
``educated'' speculations that deserve serious consideration and they
may help clarifying the complicated behavior of cuprates in the underdoped
regime. Note, once again, that the readers are encouraged to consult the 
literature presented here (e.g. \cite{ref3} and \cite{ref4}) as well
as the rest of these proceedings, to find other papers with analogous ideas.
Percolative concepts in cuprates have been around for some time, although
they have not been at the forefront of the theoretical developments.

The potential similarities cuprates-manganites lead to the
following possibilities:

* The phase diagram of $\LSCO$ in one inset of Fig.~8 is believed to contain
a spin-glass phase. From the main result of the same figure, this
phase could instead arise from a mixture of SC and AF clusters. The different
orientations of the order parameters in different clusters
could lead to their global cancellation (the order parameter
for SC contains a phase factor that could randomly change from cluster to
cluster). The implication, then, is that the phase transition SC-AF in the 
clean limit with no sources of disorder could have \emph{first-order}
characteristics, as sketched in Fig.~14 (a). This is reminiscent of
the behavior in organic superconductors and in $SO(5)$ theories of 
superconductivity \cite{ref15,ref16}.


\begin{figure}[ht]
\centerline{\includegraphics[width=.85\textwidth]
{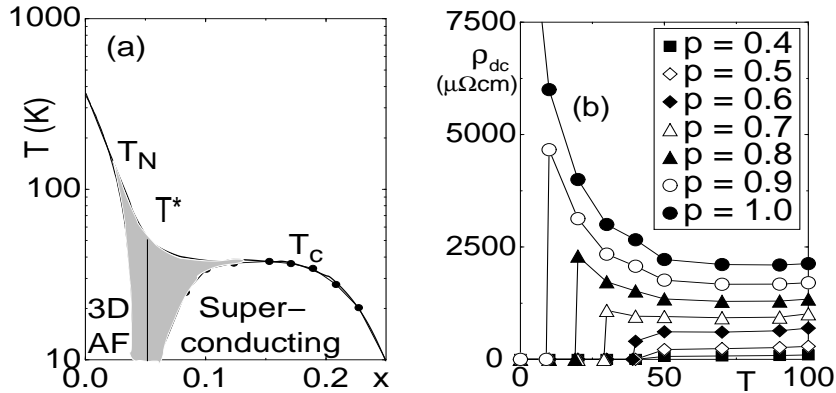}}
\caption[]{
(a) Conjectured phase diagram for high-temperature superconductivity
\cite{ref4}.
Black lines should be the actual phase boundaries without
disorder. The shaded region is conjectured to have metallic (SC) and
insulating (AF) coexisting regions in the real materials.
(b) Resistivity $\rho_{ab}$
vs T, from a random-resistor network calculation as
in Ref.\cite{ref17}, where details can be found. A 50$\times$50
cluster was used, with $\rho_{ab}$ for insulating (optimal doping)
fraction p=1.0 (0.0)
taken from LSCO x=0.04 (0.15) data [H. Takagi {\it et al.},
Phys. Rev. Lett. {\bf 69}, 2975 (1992). See also
Y. Ando {\it et al.}, cond-mat/0104163]. The inset labels
are the p fractions at 100 K, all of which are smoothly reduced
with decreasing T
until percolation to a SC state occurs at p=0.5.
}
\end{figure}

* The famous pseudogap temperature
scale would be in this context just the Griffiths temperature of
Fig.~8, namely, a remnant of the clean limit phase transition below
which the system orders locally. The results of Iguchi {\it et al.}
\cite{ref6} with superconducting regions in $\LSCO$ even at temperatures
as high as 80~K, suggest that indeed $\Tc$ could be much higher than
previously believed in this context.

* Percolative effects may exist in cuprates as well, as they seem
to be present in manganites. This is a natural consequence of the 
inhomogeneous picture for the underdoped regime \cite{mello}.
Some rough calculations
of resistivities using random resistor networks have been already 
presented (see Fig.~14 (b), reproduced from \cite{ref4},
and also Ref. \cite{ref17}).

* Although not discussed in detail in this paper, studies in Mn-oxides
\cite{ref18} have brought forward the ideas of Imry and Ma about phase
competition using the Random Field Ising Model \cite{ref19}.
In this context, when a cluster of radius $R$ is created
inside  a region dominated by a competing phase, near a first-order transition,
the energy penalization due to the surface is positive and grows
as $R^{d-1}$ where $d$ is the dimension. On the other hand, the
fluctuations in the disorder strength grow as $R^{d/2}$, from standard
considerations involving random numbers. A balance between the two
occurs at $d$=2, widely believed to be the critical dimension of the
problem. This dimension is of relevance for two-dimensional 
cuprates.  Effects of this nature could
be at work in bilayered manganites as well.
If Heisenberg
variables are used instead of Ising variables, it can be shown that
the critical dimension increases to 4, and the ideas may even apply in
three dimensional systems.

* If it is correct that cuprates and manganites are both described
by the same phenomenological approach contained in Fig.~8, then
there has to be an analog in Cu-oxides of the ``colossal'' MR
in Mn-oxides! From the discussion in previous sections it seems that
CMR occurs when preformed FM clusters are rapidly aligned by an
external magnetic field. In the SC vs. AF case, superconducting
clusters may be preformed with a basically random phase. If one
could have an external field that favors the alignment of these
phases, this theory predicts that rapidly a SC state should be
generated. However, there is no external field that we know that
can produce this alignment. An approximation would be to
bring the sample in the clustered regime in close proximity to
a system already superconducting. This would favor the phase
alignment. In fact, there are already results in the literature
reporting a ``colossal proximity effect'' in YBCO, that may be
a manifestation of the theory discussed here \cite{ref19p5}. This is an
exciting area of research that may lead to many surprises.

* Disorder may play a role in cuprates far more important than
previously anticipated. By carefully growing samples with as little disorder
as possible, the $\Tc$ should grow, having $T^*$ as the
best value possible. Purely phenomenological
studies in this context by Attfield {\it et al.} \cite{ref20},
both for manganites and cuprates, lead to similar conclusions.
Recent results by Eisaki {\it et al.} \cite{ref21} also suggest that
carefully prepared samples have higher critical temperatures than 
previously believed. Could it be that a new generation of ultra-clean
samples is needed to make progress in high temperature superconductors?
This is not a pleasant thought,
but we may need even better crystals than currently
available to unveil the proper phase diagram of Cu-oxides.

* Finally, note that materials such as $\rm CeCoIn_5$
heavy-fermions appear to
have a phase diagram quite similar to those of cuprates, including
a pseudogap regime and a $T^*$ \cite{ref22}. We feel that it is unlikely that 
totally different mechanism are at work in these
families of compounds. Thus,
the explanation for pseudogap and $T^*$ must be simple and general,
and the one described above satisfies these requirements. We do not
believe that exotic two dimensional
states are responsible for these features.

It is important to remark that there are alternative viewpoints
to the hypothesis of a first-order transition
smeared by quenched disorder into a mixed-phase state discussed here. The
alternative is to have in the clean limit
{\it tetracritical} behavior in the competition
between the antiferromagnet and the $d$-wave 
superconductor \cite{sachdev,calabrese} (as opposed
to the bicritical or tricritical behavior addressed here). This leads
to a coexistence of the two order parameters which occurs even locally,
contrary to the view postulated here and in manganites that there is
a microscopic separation of phases in real space. Some experiments 
support this view \cite{kastner}. We have not studied
how quenched disorder influences on a clean-limit tetracritical phase
diagram. Hopefully this issue will be clarified in future work.


\section{CONCLUSIONS}

In recent years the key role of inhomogeneities in transition metal
oxides and related compounds has been unveiled. The evidence in manganites
is very strong, both in theory and experiments. The competing phases
here are CO/AF and FM. The existence of preformed clusters and its easy
alignment with modest magnetic fields leads to
a large magnetoresistance. Phase separation appears to be at the heart of this
phenomenon. By mere analogy with the Mn-oxide phenomenology, speculations
can be made for cuprates as well. The spin-glass phase could arise from
SC vs. AF phase competition, and the pseudogap $T^*$ could be
a Griffiths temperature where local clusters start forming. Colossal effects
could be present in Cu-oxides, a challenging concept. The importance of
inhomogeneities in cuprates is slowly being unveiled by experiments,
and these notorious deviations from an homogeneous state must be considered
in any serious theoretical description of the still
poorly understood high temperature superconductors.

This work was supported by NSF grant DMR-0122523 and by MARTECH.

Conversations with A. Millis and S. Sachdev are gratefully
acknowledged.


\end{document}